# Language and Temporal Aspects: A Qualitative Study on Trigger Interpretation in Trigger-Action Rules

Margherita Andrao [1,2][0000-0003-2245-9835], Barbara Treccani [1][0000-0001-8028-0708] and Massimo Zancanaro [1,2][0000-0002-1554-5703]

[1] University of Trento, Trento, Italy
[2] Fondazione Bruno Kessler, Trento, Italy
margherita.andrao@unitn.it, barbara.treccani@unitn.it, massimo.zancanaro@unitn.it

**Abstract.** This paper presents a qualitative study that investigates the effects of some language choices in expressing the trigger part of a trigger-action rule on the users' mental models. Specifically, we explored how 11 non-programmer participants articulated the definition of trigger-action rules in different contexts by choosing among alternative conjunctions, verbal structures, and order of primitives. Our study shed some new light on how lexical choices influence the users' mental models in End-User Development tasks. Specifically, the conjunction *"as soon as"* clearly supports the idea of instantaneousness, and the conjunction *"while"* the idea of protractedness of an event; the most commonly used *"if"* and *"when"*, instead, are prone to create ambiguity in the mental representation of events. The order of rule elements helps participants to construct accurate mental models. Usually, individuals are facilitated in comprehension when the trigger is displayed at the beginning of the rule, even though sometimes the reverse order (with the action first) is preferred as it conveys the central element of the rule.

## 1 Introduction

End-User Programming aims to enable naive users to create programs to automate the behavior of their digital artifacts [1, 2]. Among different possible solutions [3], Trigger-Action Programming (TAP) is an event-based paradigm in which users can create rules for associating a specific trigger with a particular action to automate the behavior of both hardware and software artifacts [4-7]. These rules are usually expressed in the form of *If* <trigger> *Then* <action> in research prototypes and in popular automation platforms, such as IFTTT and Zapier [8].

Although quite easy to understand and commercially successful, this approach has the limitation of being well suited just for automating easy tasks by expressing simple rules with one trigger and without limiting or defining conditions [7]. Previous works focus, among other aspects, on the distinction between the notion of events and states as a way to facilitate the expression of more elaborated conditions in the if part of the rule [6, 9] suggesting to exploit the lexical difference between "*when*" and "*while*" to differentiate between two aspects. In addition, other linguistic aspects may support people to create different mental representations of states and events. In some languages, for example, a different verbal structure may be more appropriate when describing states rather than events. In Italian, the progressive periphrases might more accurately describe a state (e.g., "*sta piovendo*" which can be translated in English as "*it is raining*"), while a verbal structure that includes the verb "*start*" or "*begin*" emphasizing its punctuality might more accurately describe an event (e.g., "*inizia a piovere*" which can be translated in English as "*it starts to rain*").

In this work, we further elaborate on the idea of exploiting lexical choices by analyzing a larger set of linguistic features such as alternative conjunctions, verbal structures, and order of primitives. In a controlled study involving 11 participants with no experience in programming, we systematically investigated how the lexical choices impact the participants' mental model of the computational machine, that is, how they imagine a machine could interpret their instructions. The present study employs a thinking-aloud protocol [10] and a qualitative analysis [11] to get a full and rich understanding of the participants' mental processes. It is meant as a first step toward a more robust study which also measures the effectiveness of the proposed solutions proposed.

Our results, although still preliminary, suggest that all three different lexical choices investigated (conjunctions, verbal form, and order) contribute to determining the users' mental model, and some combinations of choices foster a proper model while others prompt ambiguity and potential errors.

The lesson learned from our study would suggest that, in order to improve TAP, it is important to offer users lexically accurate interfaces but also it is important to align the ontology (the representation of the domain) of the system to fully and properly represent the distinction between events and states.

### 1.1 Related work

TAP is a widely used approach due to its simplicity and intuitiveness, perfectly suitable for people without programming experience [7, 12, 13]. However, the simplicity of this model is also its limitation. For example, IFTTT only allows the creation of basic rules with a single event as a trigger restricting the expressiveness of the programs that users can create [6, 7]. In their study, Ur and coll. [7] analyzed a collection of 1590 trigger-action programs in the domain of Smart homes. Their analysis showed that 77.9% of program behaviors fitted in a single trigger and a single action form, but 16.9% required multiple triggers and possibly multiple actions. People need more expressiveness in trigger composition than those provided by common single-trigger and single-action rules to enable effective programming. Several works on TAP noted this limitation and proposed systems that allow conjunctions of multiple triggers [4, 6, 7, 12, 14]. Trigger conjunctions are meaningful for combining one event trigger alongside multiple conditions [13], similar to ECA (Event-Condition-Action) rules used by expert programmers. These studies showed that users successfully write programs with multiple triggers and actions regardless of prior experience [4, 13, 15].

However, dealing with multiple triggering conditions may be problematic for naive users. According to some results [5, 6], one of the possible causes of ambiguities and errors in rules composition, interpretation, and debugging could be an incomplete or incorrect mental model. Understanding and guiding users' mental models is crucial in Human-Computer Interaction [16, 17]. Indeed, the primary source of confusion in interacting with an artifact is due to users having wrong or inaccurate mental models of the actual functioning of the system.

Mental models are internal structures representing declarative, procedural, and inferential knowledge about the world as well as information extracted by perceptual processes [18]. When interacting with a given system, users rely on their representation of how the system works, its structure, and operations (i.e., conceptual model) stored in long-term memory [17, 18]. The combination of that information and the one extracted by perceptual processes creates a mental representation of the real interaction in working memory [19, 20]. Mental models built in working memory are, by nature, dynamic representations that allow people to simulate the piece of the world they are interacting with, supporting comprehension and reasoning processes, and allowing outcomes prediction of different scenarios. Mental models play a central role in human cognition, and (the creation of) faulty mental models are responsible for most errors in thinking (e.g., [9]).

Several studies in psychology of programming and EUD have focused on different factors involved in users' mental model creation. According to Norman [16, 17], effective conceptual models of a system should be implicitly induced (i.e., without explicit instructions) by the system design. However, information and instructions provided to users may be crucial as the systems' complexity increases. Studies investigating the effect of different descriptions on participants' performance suggest that more detailed (albeit more complex) information about system functioning allows users to develop better conceptual models. This results in more effective, as well as more satisfactory, interactions with their digital device and more accurate outcomes predictions (e.g., [21–24]). Moreover, naive users' mental models seem to be influenced by specific properties of the language and structures of rules used for EUP tasks [25, 27]. According to Pane et al. [28], the logical structure of TAP best corresponds to the natural way naive users express rules. Indeed, the authors found that the majority of the statements that non-expert participants produced spontaneously during a programming task started with "*if*" or "*while*".

Dealing with trigger conjunctions led to questions about semantics, temporal features, possible combinations of different types of occurrences, and how these affect users' mental models [5, 6, 21, 22, 29]. Huang and Cakmak [6] emphasized the role of trigger temporal features in the comprehension, interpretation, and composition of trigger-action rules by naive users. The authors defined a minimum of two types of temporal feature-based elements that can be used in triggers: instantaneous events (or simply events) and protracted over time ones (also called states) which are conditions evaluated as true or false at any time. They also observed that users often confused events and states. More specifically, the lack of distinction between different trigger types in the *If* <trigger> *Then* <action> metaphor may create ambiguities (e.g., the interpretation of when exactly triggers will occur) and undesirable outcomes, espe-cially in the context of trigger conjunctions. The authors thus suggested the following solutions at the interface level that could improve users' mental models: (i) grouping or clearly naming states and events; (ii) using different temporal conjunctions for events (e.g., "*when*", "*if* ") and states (e.g., "*while*", "*as long as*"); (iii) employing different verbs to support events (active verbs such as "*turns*") and states (present tense of the verb "*be*" such as "*I am currently at*"). Overall, they emphasize the need to investigate and implement strategies to communicate to users a clear and categorical distinction between temporal feature-based triggers to facilitate the creation of effective mental models [5, 6, 30].

The distinction between these different types of occurrences is supported implicitly by language [31, 32] grounded in the semantics and codified in lexical choices. When two simultaneous occurrences are present in the same sentence (e.g., states and events), one of them is perceived as the main one (i.e., the figure), while the other one is interpreted as the ground (i.e., the context) [33]. In particular, the longer occurrence is often perceived as the ground, and it is considered more acceptable by participants if introduced by "*while*" instead of "*when*" [33]. Furthermore, in more recent studies, the targeted use of temporal conjunctions (i.e., "*while*" to introduce states, and "*when*" to introduce events) was used to convey a distinction between states and events within trigger-action rules [21, 22, 34]. The results showed that this distinction helped participants create a clearer differentiation between the two kinds of occurrences.

Another language-related aspect that might influence mental representations is the syntactic order of temporal sentences [33]. Indeed, the iconicity assumption [35] states that by default, readers/listeners assume that the order of clauses in a sentence corresponds to the actual order of events [36]. In the case of simultaneous events (introduced by temporal conjunction such as "*while*" / "*when*"), representations are always more difficult since it violates the iconicity assumption. De Vega and coll. [33] found that, in multi-clauses sentences, the temporal conjunction indicating simultaneity ("*while*" or "*when*") is more easily understood by participants when it is at the beginning of the sentence (vs. embedded - at the beginning of the second clause). For example, the sentence "*while you are cooking, the doorbell rings*" is more easily represented in mental models than "*the doorbell rings while you are cooking*" since the first immediately informs of the simultaneity of the two events avoiding people's working memory overload.

Accumulated evidence strongly indicates that linguistic aspects need to be taken into account when designing an interface and choosing the most appropriate verbal primitives in a TAP system. Blackwell [37] suggested that many naive users' difficulties arise from the lack of direct manipulation of the elements involved in the rules and the use of abstract notational elements. Many bugs and difficulties in EUP may arise from the excessive/unnecessary distance between users' mental models and the adopted programming languages [38], together with users' inclination to transfer language knowledge into programming tasks [39].

# 2 The study

The study aimed to investigate the role of some specific aspects of language in trigger-action rules composition and interpretation. In particular, we explored the influence of different temporal conjunctions ("*as soon as*", "*when*", "*if*", "*while*", and "*as long as*") and verbal structures (progressive periphrases which are analogous to the present progressive tense in English such as "*it is raining*", specific marks to signal instantaneousness such as "*it starts to rain*", compared to generic forms in present indicative such as "*it rains*") to facilitate the distinct representation of states and events [6]. A second aspect under investigation was the study of the influence of the syntactic order with which trigger and action were presented [33].

The study was organized as individual sessions in which we asked participants, without programming experience, to read a few scenarios illustrating desired states and to compose TA rules aimed at achieving them. The scenarios were carefully prepared to compare different combinations of these variables.

The scenarios were presented one at a time and, after reading a scenario description, the participant had to select the language primitives to compose the trigger part of the rule and decide the preferred order of the trigger and the action part. A thinking-aloud technique [10] was used to collect verbal reports on participants' mental models.

## 2.1 Participants

Seven females and five males, aged between 21 to 33 years (M = 27.67 and SD = 3.49 years; Mf = 27.43 and SDf = 3.41 years; Mm= 28.00 and SDm=3.41 years), participated in the study. All of them were native Italian speakers with no experience in programming. Three of them were non-degree workers, while nine were university students or just-graduated workers. Participants were recruited by means of a snowball sampling in the surrounding areas of Trento and Brescia (in the north of Italy). Data related to one female participant was excluded due to technical problems during the recording.

## 2.2 Materials

We created 24 impersonal descriptions of scenarios describing everyday life situations (see Table 1). In half of the scenarios, the desiderated outcome could be achieved with a rule triggered by an event and, in the other half, with a rule that should be maintained active during a state. We identified 12 events (for example, "*to rain*") that fit both a state-based, protracted, interpretation ("*it is raining*") and an event-based, instantaneous, interpretation ("*it starts raining*"). For each occurrence (event or state), we defined three verbal structures: an event-specific (preceded by "*it starts to*"), state-specific (progressive form), and generic form (present tense). Then, we defined an associated action for each scenario (e.g., "*keep the umbrella open*" for the state-based scenario, "*close the window immediately*" for the event-based one).

All material was designed and presented in Italian. Figure 1 summarizes all 24 scenarios.

**Table 1.** In the table is visible the list of 24 scenarios used in our study and the language primitives (original Italian version and English translation) describing actions and triggers.

| Scenario | | | Language Primitives | | | |
|---|---|---|---|---|---|---|
| n | occurence | description | action | verb generic | verb state-specific | verb event-specific |
| 1 | State | Avoiding clothes getting wet | *tieni l'ombrello aperto*<br>keep the umbrella open | *piove*<br>it rains | *sta piovendo*<br>it is raining | *inizia a piovere*<br>it starts to rain |
| 2 | Event | Avoiding water entering the house | *chiudi subito la finestra*<br>close the window immediately | | | |
| 3 | State | Increasing the stimuli for waking up | *tieni alto il volume*<br>keep the volume up | *la sveglia suona*<br>the alarm clock rings | *la sveglia sta suonando*<br>the alarm clock is ringing | *la sveglia inizia a suonare*<br>the alarm clock starts to ring |
| 4 | Event | Having a coffee ready in the morning | *accendi subito la macchinetta del caffè*<br>turn on the coffee machine immediately | | | |
| 5 | State | Avoiding light entering the room | *tieni le tapparelle abbassate*<br>keep the blinds down | *splende il sole*<br>the sun shines | *sta splendendo il sole*<br>the sun is shining | *inizia a splendere il sole*<br>the sun starts to shine |
| 6 | Event | Hanging the laundry outside | *invia subito una notifica*<br>send notification immediately | | | |
| 7 | State | Recording Carlo's speech | *tieni il microfono acceso*<br>keep the microphone on | *Carlo parla*<br>Carlo speaks | *Calro sta parlando*<br>Carlo is speaking | *Carlo inizia a parlare*<br>Carlo starts to speak |
| 8 | Event | Listening to Carlo's speech | *abbassa subito il volume della musica*<br>turn down the music volume immediately | | | |
| 9 | State | Avoiding missing parts of the lesson | *tieni il registratore acceso*<br>keep the recorder on | *il Professore spiega*<br>the Professor teaches | *il Professore sta spiegando*<br>the Professor is teaching | *il Professore inizia a spiegare*<br>the Professor starts to teach |
| 10 | Event | Avoiding interrupting the lesson | *attiva subito la modalità silenziosa*<br>turn on silent mode immediately | | | |
| 11 | State | Working in an optimal environment | *tieni la temperatura a 18 gradi*<br>keep the temperature at 18 degrees | *si lavora*<br>one works | *si sta lavorando*<br>one is working | *si inizia a lavorare*<br>one starts to work |
| 12 | Event | Avoiding inappropriate calls | *attiva subito la modalità non disturbare*<br>activate do not disturb mode immediately | | | |
| 13 | State | Increasing concentration in study | *tieni la porta chiusa*<br>keep the door close | *si studia*<br>one studies | *si sta studiando*<br>one is studying | *si inizia a studiare*<br>one starts to study |
| 14 | Event | Reducing distractions | *disattiva subito tutte le notifiche*<br>turn off all notifications immediately | | | |
| 15 | State | Reducing background noise | *tieni attiva la cancellazione del rumore*<br>keep noise cancellation on | *il cane abbaia*<br>the dog barks | *il cane sta abbaiando*<br>the dog is barking | *il cane inizia ad abbaiare*<br>the dog starts to bark |
| 16 | Event | Creating a quiet environment | *apri subito la porta*<br>open the door immediately | | | |
| 17 | State | Reaching maximum effort at the sports medical exam | *tieni monitorata la frequenza cardiaca*<br>keep your heart rate monitored | *si corre*<br>one runs | *si sta correndo*<br>one is running | *si inizia a correre*<br>one starts to run |
| 18 | Event | Recording the number of miles of a workout | *attiva subito il conta passi*<br>activate the step counter immediately | | | |
| 19 | State | Responsible traffic circulation | *tieni i fari accesi*<br>keep headlights on | *la macchina si muove*<br>the car moves | *la macchina si sta muovendo*<br>the car is moving | *la macchina inizia a muoversi*<br>the car starts to move |
| 20 | Event | Listening to a podcast in the car | *accendi subito la radio*<br>turn on the radio immediately | | | |
| 21 | State | Alerting people of a fire | *tieni attivo l'allarme anti-incendio*<br>keep the umbrella open | *si rileva del fumo*<br>one detects smoke | *si sta rilevando del fumo*<br>one is detecting smoke | *si inizia a rilevare del fumo*<br>one starts to detect smoke |
| 22 | Event | Preventing a fire from spreading | *accendi subito gli irrigatori anti-incendio*<br>close the window immediately | | | |
| 23 | State | Avoiding household accidents | *tieni spenti i fornelli*<br>keep the stove off | *se si dorme*<br>one sleeps | *si sta dormendo*<br>one is sleeping | *si inizia a dormire*<br>one starts to sleep |
| 24 | Event | Going to bed safely | *attiva subito l'allarme*<br>activate the alarm immediately | | | |

## 2.3 Methods

Data were collected in a controlled setting using the online platform Qualtrics. Each participant answered all 24 scenarios. Every trial consisted of two different parts: the composition task and the order preference one.

In the composition task, participants were instructed to define the language primitives for creating the sentence most appropriate as instructions to realize the scenario proposed. Participants were seated in front of a computer screen. They read the scenario described on the screen. Immediately under this description the action part was presented. Two dropdown lists allowed participants to select the temporal conjunction (among five: "*as soon as*", "*when*", "*if*", "*while*", and "*as long as*") and the verbal structure (among three: generic form, state-specific, event-specific) to compose the rule (see Fig. 1).

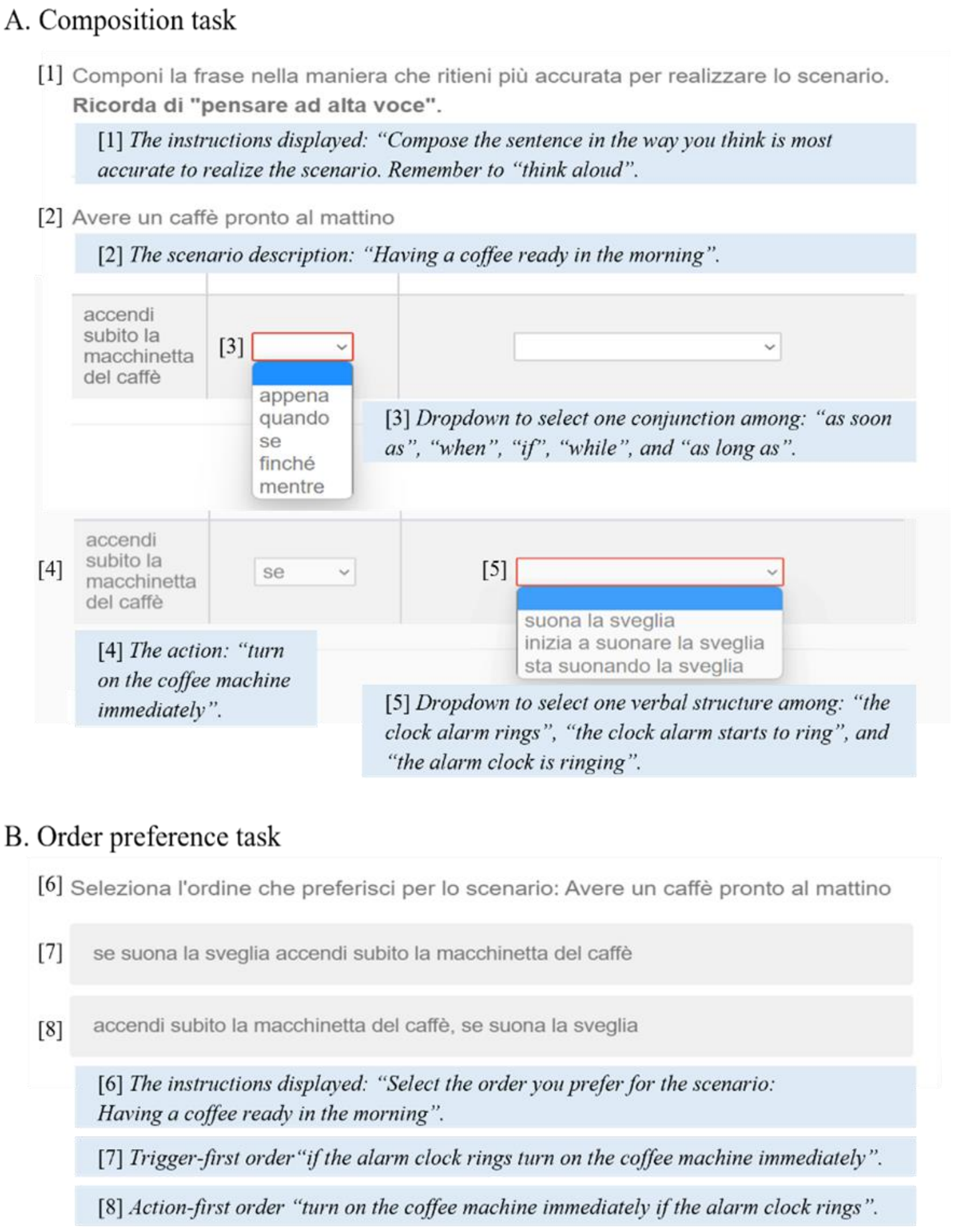

**Fig. 1.** Two s screenshots from Qualtrics platform showing the two tasks performed by the participants. (A) The top part of the figure shows an example of the composition task for the scenario "*Having a coffee ready in the morning*". (B) The bottom part of the figure represents the second task to detect the order preference for the same scenario (the light blue boxes contain English translations, and they are not part of the interface).

Subsequently, in the order preference task, participants were asked to choose the order they perceived as the most natural and accurate to compose an instruction to realize the scenario. In this second screen, participants saw the selected choice in two orders: trigger-first (e.g., "*if it rains close the window immediately*") and action-first (e.g., "*close the window immediately if it rains*"). Finally, participants filled in a short demographic form for age, gender, and previous experience with programming languages. Following a think-aloud protocol, participants were explicitly asked to motivate and elaborate their decision. Each session lasted approximately 45 minutes and was video and audio-recorded. Session recordings were transcribed and then coded and analyzed using thematic content analysis [11] by two independent researchers.

### 2.4 Results from Thematic Analysis

A thematic analysis of participants' verbal reports helped us to identify four themes around which to try to understand the relation between lexical choices and participants' mental models of the programming activity.

**Theme 1: "As soon as" effectively highlights the instantaneousness of events, especially when associated with the verbal structure "starts to".** Participants identified the conjunction "*as soon as*" as the most accurate and precise for describing the timeliness of events. This conjunction helped participants mentally represent a specific instant, as some participants explained:

*Participant 3: "'As soon as' makes me think of it as immediate because it is a gesture, an immediate activity".*
*Participant 9 [referring to scenario n. 6]: "The word 'as soon as' makes me think of an instant [...] the instant when the sun comes out".*
*Participant 10 [referring to scenario n. 4]: "'As soon as' [...] even though maybe the alarm clock rings for a few seconds, I interpret it as the first second it rings".*
*Participant 12: "As soon as this thing happens, so at that precise moment".*

The conjunction "*as soon as*" seemed to provide confidence to some participants that the action would be carried out effectively. As explicated by participant 8:

*Participant 8 [referring to scenario n. 7]: "'As soon as' gives me more confidence in having recorded everything that Carlo has to say."*

Often this conjunction was associated with the verbal structure "*starts to*" (event-specific) to reinforce the instantaneousness but also the timeliness of the occurrence. For example, some participant motivated their choices by explaining:

*Participant 5 [referring to scenario n. 18]: "because it is 'as soon as one starts to run', immediately activates the step counter [...] to strengthen the sentence".*
*Participant 8 [referring to scenario n. 6]: "I would choose 'as soon as the sun starts to shine' [...] I need to do the action as soon as the first ray of sunshine enters the room".*

For participant 11, the verbal structure "*starts to*" (event-specific) was implicitly associated with the conjunction "*as soon as*" influencing his choices:

*Participant 11 [referring to scenario n. 10] "Again, implicitly I add 'the Professor starts to teach' because there is the 'as soon as'".*

Nevertheless, for the same participants, this association felt redundant and unnecessary if other elements (e.g., the semantics of the verb) convey the same temporal information:

*Participant 11 [referring to scenario n. 4]: "Always 'as soon as' because it tells me the beginning of the action [...] here I would even put without the 'it starts to' but just 'as soon as the alarm rings' because it gives me the idea of a shorter duration this thing [...] the alarm clock lasts up to 10 seconds. If I say, 'as soon as' is enough for me to tell 'as soon as it starts to ring'".*

In general, the verbal structure "*starts to*" (event-specific) helped to represent precisely the instant when the action had to be performed. Almost all participants emphasized this aspect several times. While choosing primitives in the composition task, some participants reflected:

*Participant 2 [referring to scenario n. 8]: "in my opinion, rather than 'as soon as Carlo speaks', 'as soon as Carlo starts to speak', it is already more precise".*
*Participant 10 [referring to scenario n. 8]: "Because it is the instant when Carlo starts to speak [...]. So for me, it's instantaneous".*

Some participants related the verbal form with the need to promptly perform the action, therefore assuming an impact on the lexical choice in the trigger with the criteria of execution of the action in the TA rule:

*Participant 4 [referring to scenario n. 12]: "'One starts to work'. This is the sentence construction when you want the action immediately".*
*Participant 8 [referring to scenario n. 4]: "'It starts to ring' gives me the idea of something you have to do instantaneously".*

*Participant 9 [referring to scenario n. 10]: "In this case, because it's an action I have to do immediately, promptly, as soon as the Professor starts to teach, I have to put the silent mode. Yes, it seems that adding the verb to start gives me even more urgency."*

**Theme 2: "while" effectively highlights the protractedness of events (especially when associated with the progressive form) and "as long as" seems to be more related to the end of the action.** The conjunction that best regarded to best express the idea of continuity and duration seems to be the "*while*":

*Participant 4: "When I read 'while' for me it is always something that lasts over time".*
*Participant 9 [referring to scenario n. 23]: "It has to last the whole time I sleep. So 'while' seems the most appropriate to me".*
*Participant 11: "I would use 'while' to be even more specifically protracted".*

Similar to what emerged for "*as soon as*", "*while*" was often associated with the progressive form because it reinforced the conception of protractedness. For example, some comments from participants 3, 4, and 5:

*Participant 3 [referring to scenario n. 13]: "In this case, 'keep the door close' is more continuous to me, so 'while' is doing an action".*
*Participant 4 [referring to scenario n. 1]: "maybe 'it is raining' [works better]. It gives more the idea of continuing to do the action".*
*Participant 5 [referring to scenario n. 7]: "Carlo is speaking. [...] I could also select 'while Carlo is speaking' but 'while' and 'is speaking' give the same sense to the sentence [...] to reinforce the idea of temporality".*

Participant 12, on the other hand, preferred not to choose the progressive form and rather used the present tense (the generic form) because, in this participant's view, other elements of the sentence already included the concept of duration:

*Participant 12 [referring to scenario n. 11]: "'while one works' and not 'is working' because it seems a repetition to me because 'while' already gives me the idea of something protracted.".*

In general, many participants (2, 3, 4, 5, 8, 9, 11, 12) identified the progressive form as reinforcing the representation of a protracted occurrence. For example,

*Participant 11 [referring to scenario n. 5]: "Here, the sun shines for a duration of time, I would say 'it is shining' because it's a protracted action".*

As participant 3 pointed out, this was the verbal structure to guarantee the action would be performed even after it began:

*Participant 3 [referring to scenario n. 19]: "More than 'it starts to move', which means that after you start you can turn off the lights, 'It is moving' gives the obligation to hold them while you are moving".*

The conjunction "*as long as*" facilitated participants to represent prolonged occurrence but unlike "*while*", it seems also to suggest or emphasize the end of the action. For example,

*Participant 5 [referring to scenario n. 5]: "'as long as the sun shines'. Because 'while' is during, 'as long as' gives it an end".*
*Participant 4 [referring to scenario n. 13]: "'as long as' is convenient here. [...] 'as long as' gives the idea that until you finish studying, you have to keep the door closed".*
*Participant 7: "instead of something that is happening, 'as long as' gives me the idea that is going on in time together with the idea of the end".*

Presumably related to the semantics of the conjunction, participant 3 associated "*as long as*" with the presence of other rules:

*Participant 3 [referring to scenario n. 1]: "it could be: 'as long as it rains'. I basically assume that at one point one opened it [the umbrella] when it started raining, and this [the rule he is composing] is after that moment. Before, there was 'open the umbrella as soon as it starts to rain' and now 'keep the umbrella open as long as it rains'".*

Moreover, "*as long as*" led participants to imagine the situation in which the rule might not be triggered. Some examples:

*Participant 8 [referring to scenario n. 15]: "as soon as it finishes barking, I'll turn off the noise cancellation mode".*
*Participant 10 [referring to scenario n.13]: "'as long as', so that then I can also open the door if I'm not studying anymore".*
*Participant 12 [referring to scenario n. 1]: "'as long as it is raining', if it's not raining, you don't get wet so you can close the umbrella".*

**Theme 3: "if" and "when" are ambiguous and they do not consistently support the distinction among events with different temporal aspects; similarly, the present tense does not support the definition of temporal aspects and therefore may generate ambiguity.** Both "*if*" and "*when*" was used to define indistinctly instantaneousness and continuity of occurrences depending on the context and other temporal elements of the rule including the semantics of the verb associated with the trigger and the duration of the action. The ambiguity of the conjunction "*when*" is apparent because participants used it to express both timeliness and duration, often unaware. For example, initially, participant 2 stated:

*Participant 2 [referring to scenario n. 8]: "'when' means as soon as he starts talking".*

Then, at one point, he contradicted himself by saying:

*Participant 2 [referring to scenario n. 1]: "'when it rains' means something that is protracted in time because it is not as soon as it starts raining".*

Other participants explicitly expressed their difficulties with the use of "*when*":

*Participant 9: "I feel like I contradict myself".*
*Participant 10 [referring to scenario n. 8]: "When Carlo speaks, immediately turns down the volume of the music, for me this is tricky".*

Participants 11 and 4 were able to recognize that "*when*" does not support individuals in defining events temporally:

*Participant 11 [referring to scenario n. 4]: "when the alarm clock rings, however even 'when' is quite unclear".*
*Participant 4 [referring to scenario n. 4]: "'when the alarm clock rings'. It doesn't sound either like something that happens immediately or an ongoing thing. It's in the middle".*

Similar comments were made for the conjunction "*if*". Participant 9, who had associated "*if*" with "*as soon as*" because of its connotation of immediacy, then expressed:

*Participant 9 [referring to scenario n. 24]: "Instead, in this case, it's different. [...] because 'if' with the verb sleep sounds like something long. So the 'if' changes according to the verb that comes after."*

Other participants expressed uncertainty and difficulty related to the temporal representation of "*if*".

*Participant 3 [commented at different moments]: "the ones with the 'if' are the most tricky [...] the if already puts you in an uncertain, doubtful form [...]. In all of them [scenarios] 'if' might also work, however, without giving certainty".*
*Participant 4 [commented at different moments]: "the 'if' indicates a condition of doubt and uncertainty [...] the 'if' associates well with all scenarios [...] I never used 'if'. The 'if' is too doubtful".*

Apparently, the conjunction "*if*" did not help the mental representation of the rule because it was perceived as very general. For example, participant 10 said:

*Participant 10 [referring to scenario n. 5]: "This one is difficult. [...] 'If the sun shines' it could be even for a few hours".*

Sometimes, however, the flexibility and temporal imprecision of "*if*" is preferred in situations of uncertainty because it limits the temporal boundaries of the rule. For example,

*Participant 5 [referring to scenario n. 20]: "it could be 'if the car moves'. Because for listening to a podcast in the car, the 'if' is the option that temporally limits you the least, [...] 'if it moves', when you want you can listen to it [the podcast]".*

The present tense for the event verb did not help in assessing the temporal definition of the occurrence, as expressed by participant 8:

*Participant 8: "'one runs' [stressing on the present indicative tense] is the concept that meets the 'beginning', the 'while' and the 'end'".*

Some participants clarified how the present tense verb defines general situations. For example, participants 7 and 8 explained:

*Participant 7 [referring to scenario n. 15]: "'When it barks' not 'when it is barking' because it could bark at any time".*
*Participant 8 [referring to scenario n. 4]: "'it starts to ring' gives me the idea of something you have to do instantaneously, and 'it's ringing' gives me the idea of a period [...] 'it rings' instead is a broader concept that gives me more space.*

**Theme 4: the temporal order with the trigger at the beginning of the rule facilitates comprehension yet the most important element to communicate should be at the beginning of the sentence and it is mainly the action.** Regarding the order of the trigger and the action part, some participants expressed a preference for the order which corresponds to the actual temporal order, with the trigger at the beginning of the rule (regardless of whether it described a state or an event):

*Participant 5 [referring to scenario n. 4]: "I would always select this way [trigger-first], with the right temporal order. That first, it happens that the alarm rings, then I turn on the coffee machine".*
*Participant 10 [referring to scenario n. 9]: "Again the sentence beginning with 'as long as' [...] it makes me understand that I have to keep it on for the whole duration".*
*Participant 9 [referring to scenario n. 2]: "Because from the temporal point of view, first, it rains, and then I close the window".*

Participant 11 and 10 explained their choices considering the logical consequentiality of condition and action:

*Participant 11: "Because it gives me the indication first of the condition and then the action I have to do in this condition. I don't care to know the action if I don't know the condition under which this action should occur".*
*Participant 10 [referring to scenario n. 2]: "because it gives me more of an idea of why I have to close the window. So it makes you understand immediately".*

Nevertheless, from the words of the participants emerged that the elements of the sentence should be ordered by importance. Some participants argued that the trigger was the crucial part since its absence implies that the action won't be performed, as participant 4 expressed:

*Participant 4 [referring to scenario n. 2]: "because it gives you the indication, 'as soon as it starts to rain', and then close the window. It seems more important to me to say the first part."*

But more participants claimed that the action should be placed at the beginning as it was the main component. Some extracts from verbal reports, as an example:

*Participant 7: "The important thing is the action".*
*Participant 8: "The main diktat [i.e., order or imposition] is 'close the window'. And then there is the part of explaining why. But in the meantime, it [the action] directs you to do something".*
*Participant 10 [referring to scenario n. 11]: "This one [the action-first order] because it gives us a better understanding of what you need to do, which is to 'keep the temperature at 18 degrees'. Then, it is less important to know when, while you are working or studying".*
*Participant 12 [referring to scenario n. 20]: "But do I care more about the car moving or listening to the radio? Listening to the radio is the thing that interests me. [...] Because normally when people listen to sentences, they lose interest at the end".*

In particular, the urgency to start the rule with the action was made explicit for scenarios related to situations of potential danger or harm to safety (fire [scenario n. 21 and 22], road traffic [scenario n. 19], home security [scenario n. 23]). In these cases, almost all participants preferred to place the action at the beginning of the rule rather than the trigger. In this case, the action (placed precisely at the beginning of the sentence) is perceived as the figure, while the trigger remains in the background, as reported by:

*Participant 3 [referring to scenario n. 19]: "In this case, however, I would put 'keep the headlights on' first because it seems more important to me [...] the action of keeping the headlights on that [...] is always a safety issue".*
*Participant 9 [referring to scenario n. 21]: "I would imagine that the alarm siren should sound during the entire period of smoke detection. In that case, I would*

*change my mind [in the other scenarios she preferred the trigger-first order] because the first part of keeping the fire alarm on seems more important to me".*

## 3 Discussion

From our results, it emerges, as already observed by Huang and Cakmak [6], that the use of different temporal conjunctions for events (e.g., "*when*", "*if*") and states (e.g., "*while"*, "*as long as*") facilitated non-programmers in effectively distinguishing occurrences with different temporal features.

The conjunction "*if*" although commonly used to define the event in trigger-action rules [8], creates ambiguity, as other studies already highlighted [21, 22, 40]. Indeed, also from our results emerges that individuals perceived "*if*" as ambiguous and they often associated it indistinctly with both events and states.

Recent studies [21, 22, 34] have suggested the use of "*when*" and "*while*" to define events and states, respectively. Our results support the idea that "*while*" is easily associated with prolonged occurrences and this association was stronger when the occurrence was described with the progressive form.

Contrastingly, it appears that "*when*" is not the conjunction that best fits the representation of instantaneous events. Indeed, our results seem to suggest that "*when*" is perceived as imprecise and does not support the temporal definition of events, similarly as"*if*". The conjunction "*as soon as*" seems to be more appropriate to define punctual events as it spontaneously conveys the idea of an instantaneous occurrence, even more, if associated with the verbal structure "*starts to*".

Even the conjunction "*as long as*" seems to have a precise and unambiguous representation that leads individuals to mentally represent the end of the event. While in some situations, this focus on the end of an action might be beneficial, it is worth noting that it may also raise unexpected errors.

Similar to the results of other studies [28], our findings support the idea that individuals were likely to prefer the order with the trigger at the beginning of the rule supporting the iconicity assumption, i.e., that the order of elements in a sentence corresponds to the actual order of events and that the conjunction placed at the beginning of the sentence facilitates the understanding of the rule [33].

In addition, we found that individuals preferred, in some specific circumstances dealing with personal safety, to reverse the order of the elements when the triggered action is considered the most important part.

The lesson learned from our study might be that, in order to improve TAP, it is important to offer users lexically accurate interfaces by favoring *"as soon as"* and *"while"/"as long as"* conjunctions, possibly aligned with redundant verbal forms and a lexical order that focuses the attention on the most relevant aspects of the rule. From an engineering point of view, still, it is also important to align the ontology (the representation of the domain) of the system to fully and properly represent the distinction between events and states. That is, the internal representation of the system should explicitly represent instantaneous events as different from protracted ones even when, as it might be often the case, the difference between the two is merely a matter of representation: for example, because the same sensors can recognize the state (for example, “*it rains*”) and the event (“*it starts to rain*”) is implicitly derived.

## 4 Conclusion

In this paper, we presented an explorative qualitative study that investigates the effects of some language choices in expressing the trigger part of a trigger-action rule on the users’ mental models.

Although the results need to be validated in a larger study, they seem to suggest that lexical choices play an important role in determining the mental model of naive users engaged in EUD tasks and some combinations of choices foster a proper model while others prompt ambiguity and potential errors. A lesson learned for EUD system engineering would be that users can be facilitated in understanding the semantics of a rule-based system by carefully crafting the ontological definition of the domain (that is, providing an internal representation of instantaneous events paired with the corresponding states, or protracted events, as separate entities) and then properly mapping the domain representation to lexical choices that reduce ambiguity for the naive user.

This study has some limitations, in particular, the limited number of participants and the focus on a specific language (although the linguistic phenomena investigated are not unique characteristics of Italian). Furthermore, other linguistic aspects might play related roles (verb aspect, among others) and different and more variegated scenarios should be investigated. Still, we believe that this study may contribute to an ongoing discussion about user-centered design of EUD systems.

**Acknowledgments** This work has been supported by the Italian Ministry of Education, University and Research (MIUR) under grant PRIN 2017 "EMPATHY: EMpowering People in deAling with internet of THings ecosYstems" (Progetti di Rilevante Interesse Nazionale – Bando 2017, Grant 2017MX9T7H). We thank Elia Baccaro for his work in recruiting and testing participants of the study as part of his internship.